# A software parallel programming approach to FPGA-accelerated computing

Ruediger Willenberg, Paul Chow

*Electrical & Computer Engineering, University of Toronto*
*Toronto, Ontario, Canada*
willenbe@eecg.toronto.edu
pc@eecg.toronto.edu

*Abstract*—This paper introduces an effort to incorporate reconfigurable logic (FPGA) components into a software programming model. For this purpose, we have implemented a hardware engine for remote memory communication between hardware computation nodes and CPUs. The hardware engine is compatible with the API of GASNet, a popular communication library used for parallel computing applications. We have further implemented our own x86 and ARMv7 software versions of the GASNet Core API, enabling us to write distributed applications with software and hardware GASNet components transparently communicating with each other.

## I. MOTIVATION

The use of FPGA accelerators by software programmers is inhibited by complexity both on the micro- and macroscopic levels. At the microscopic level, an efficient computation needs to be implemented in a completely different way than in software code, considering dataflow and pipelining aspects of customized hardware. While most of this work is currently done in low-level industry standard languages like VHDL and Verilog, more promising modern languages and High-Level-Synthesis (or "C-to-Gates") tools promise easier design paths for the future.

On the macroscopic or system-design level, every FPGA and platform vendor uses different toolchains, CPU-FPGA interfaces and FPGA-memory interfaces. In contrast to the software side, where standard languages and operating system APIs facilitate the writing of portable code, every change of FPGA platform involves major design modifications. FPGA vendors are currently introducing OpenCL[1] support to allow the use of FPGA resources through an industry standard interface originally established for programming GPUs. In our opinion, this approach overconstrains the design flexibility of FPGAs by imposing memory hierarchy and execution models from the GPU domain. However, it correctly determines that, on the macroscopic level, the communication and synchronization patterns of data and control flow do not differ markedly between software and hardware components. In our work, we are exploiting this observation by offering a common API for both hardware and software modules.

## II. UNIFIED PROGRAMMING MODEL AND API

In our opinion, a unified programming model and API for all components in a heterogeneous system, as shown in Figure 1,

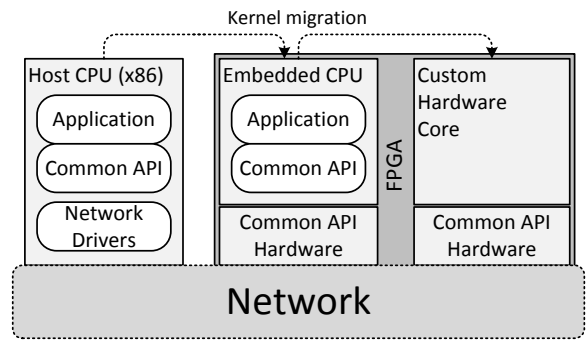

Fig. 1. Example for a multiple-platform system with unified parallel API: Host CPU, embedded CPU on FPGA and custom FPGA hardware

is beneficial to keeping applications both maintainable and scalable. In this approach, the same API is used by both application processes running on host CPUs as well as on embedded CPUs located on the FPGA. The latter can be either *soft processors* implemented in the logic fabric or hardwired cores. Furthermore, custom hardware components will use a similar "API" by using the same set of configuration parameters to control synchronization and data communication. Instead of FPGA components just being utilized as co-processors, hardware and software components are treated as equal interacting processes.

A significant benefit of a common API is the easy migration of performance-critical application kernels to hardware. An algorithm can be implemented, verified and profiled completely in software, taking advantage of the sophisticated development and debugging tools available in this domain. When profiling has identified the code sections taking up the most execution time, these can be converted into custom hardware cores and seamlessly integrated into the original software component architecture.

Previous work has successfully demonstrated this approach using the Message Passing Interface [2]. Our current work implements the same approach for the GASNet (Global Address Space Networking) API[3], which allows applications direct access to all shared memory regions in a parallel

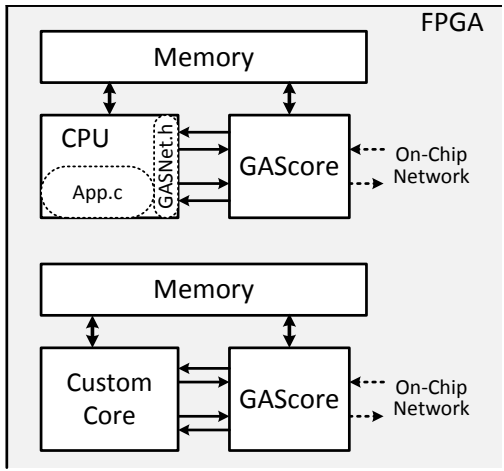

Fig. 2. GAScore configurations with CPU and custom hardware core

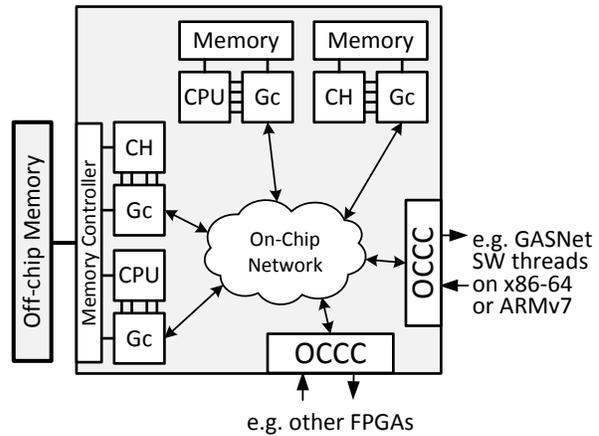

Fig. 3. FPGA with embedded CPUs, custom hardware cores (CH), GAScores (Gc), on-chip and off-chip RAM, and possible off-chip connections (Off-Chip Communication Controller, OCCC)

system, including the disparate memory architectures found in a heterogeneous system. GASNet uses the concept of *Active Messages*, network packets that include a data payload, a destination address and the identity of an asynchronous *handler function* to call after receipt of the data.

## III. HARDWARE API SUPPORT

### A. Remote DMA via GAScore

To allow computation nodes one-sided communication to and from remote memories, we have built a hardware component called *GAScore* (Global Address Space core). Its main function is to act as a Remote DMA controller that can read and write data in another computation node. GAScore command sequences use the same arguments as function calls to GASNet. Data packets sent between GAScores are *Active Messages*.

Each GAScore is connected to the same working memory used by its local CPU or computation core as shown in Figure 2. An on-chip CPU or a custom hardware core initiates an Active Message by sending a command to GAScore through a FIFO hardware connection. To execute a remote write, GAScore reads the local data and sends it off through an on-chip network. On the receiving side, a GAScore or GASNet software thread writes the data from the message packet into the destination memory and calls the local handler function on the CPU or hardware core.

### B. On-chip infrastructure

A single FPGA can hold multiple GASNet processing nodes as illustrated in Figure 3. Both embedded CPUs and custom hardware cores can execute different algorithm segments or work on different data sets concurrently. They are connected through the GAScores to a simple packet-based on-chip network. Off-Chip Communication Controllers (OCCC) connected to the on-chip network can receive and transmit messages from and to host buses, local area networks and custom inter-FPGA or inter-board connnections. For the computation nodes, network addressing is limited to sending Active Messages to other GASNet nodes. Each individual node can work with either a local on-chip memory or shared off-chip memory, depending on the type and size of dataset being processed.

## IV. PROJECT STATUS

The described supporting infrastructure for soft processors and custom hardware processing cores has been successfully implemented on four-chip FPGA platforms based on Xilinx Virtex-5 and -6 devices. Basic performance and overhead results for these FPGA-only systems have been reported in an earlier publication[4].

We have recently completed software GASNet implementations that can connect to the described FPGA infrastructure by PCIe(x86) or AXI(ARM). GASNet nodes in x86 or ARM software threads, MicroBlaze software nodes and custom hardware nodes can successfully communicate with each other through the common API. We are currently implementing benchmarking applications to demonstrate the performance of heterogeneous GASNet-based systems.

We are further planning higher-level communication abstractions in the form of a C++ library for parallel scientific computations.